\documentclass[twocolumn,a4paper,amsmath,amssymb,showpacs,prl,superscriptaddress]{revtex4}

\usepackage{graphicx}
\usepackage{epstopdf}

\begin{document}

\title{First-Principles Theory of Multipolar Order in Neptunium Dioxide}

\author{M.-T. Suzuki}
\affiliation{Department of Physics and Astronomy, Uppsala University, Box 516, S-751 20 Uppsala, Sweden}
\author{N. Magnani}
\altaffiliation{Current address: Actinide Chemistry Group, Chemical Sciences Division, Lawrence Berkeley National Laboratory, 1 Cyclotron Road Mail Stop 70A1150, Berkeley CA 94720-8175, USA.}
\affiliation{European Commission, Joint Research Centre, Institute for Transuranium Elements, \\ Postfach 2340, D-76125 Karlsruhe, Germany}
\author{P. M. Oppeneer}
\affiliation{Department of Physics and Astronomy, Uppsala University, Box 516, S-751 20 Uppsala, Sweden}

\date{May 4, 2010}

\begin{abstract}
We provide a first-principle, materials-specific theory of multipolar order and superexchange 
in NpO$_2$ by means of a non-collinear local-density approximation +$U$ (LDA+$U$) method.
Our calculations offer a precise microscopic description of the triple-$\mbox{\boldmath$q$}$-antiferro ordered phase in the absence of any dipolar moment. We find that, while the most common non-dipolar degrees of freedom (e.g., electric quadrupoles and magnetic octupoles) are active in the ordered phase, both the usually neglected higher-order multipoles (electric hexadecapoles and magnetic triakontadipoles) have at least an equally significant effect.
\end{abstract}

\pacs{71.20.-b; 75.10.-b; 75.30.Et}

\maketitle

One of the milestone achievements in the field of magnetism was the discovery that the relatively simple Heisenberg-Dirac exchange Hamiltonian successfully describes the intersite interactions for most $3d$-electron-based compounds \cite{anderson}.
In contrast, superexchange interaction involving terms other than magnetic dipoles can become effective when the orbital degrees of freedom are unquenched and give rise to exceptional physical behaviour, such as unconventional ordered phases governed by
multipolar order parameters \cite{santini09,kuramoto09}.
Unfortunately, in most cases recognizing the true driving force behind these anomalous phase transitions is a complicated task, due to the large number of independent degrees of freedom to be considered. For example, more than half a century of extensive experimental and theoretical studies \cite{westrum53, ross67, cox67, caciuffo87, dunlap68, friedt85, mannix99, kopmann98, santini00, paixao02, tokunaga05, kiss03, kubo05a, magnani08, magnani05} was required to reach a consensus on the real nature \cite{santini06} of the unusual ordered state displayed by neptunium dioxide below 25 K. NpO$_2$ is now widely recognized as the archetypal compound for magnetic multipole ordering \cite{santini09}.
Part of the problem is due to the fact that, although crystal-field theories (and their extensions) provided precious insight in the local symmetry of the $5f$ levels and the possible order parameters \cite{santini00,kiss03,kubo05a,santini06,kuramoto09}, these do not include any material-specific aspects, such as hybridization with the O $2p$ electrons, and cannot explain why, of all materials, NpO$_2$ is special;
also, models involving a cumbersome large number 
of adjustable parameters would have to be used if all active degrees of freedom were to be considered; even the most recent theoretical advances \cite{kubo05a,santini06,santini09} have to rely on drastic assumptions \cite{shiina97} to simplify the calculations, notwithstanding the wealth of available experimental data.

In this Letter we provide a first-principles theory of multipolar order in NpO$_2$.
We show that non-collinear local density approximation +$U$ (LDA+$U$) calculations can successfully overcome the above-mentioned difficulties and provide new insight in the anomalous symmetry-broken phase.
Somewhat surprisingly, only a few first-principles studies \cite{maehira07,prodan07,wang10} were ever performed on NpO$_2$, none of these concerning multipolar ordering because the necessary computational framework was not developed.
Conversely, we show here that our state-of-the-art full-potential density-functional theory based calculations, using the LDA$+U$ method \cite{anisimov91} to treat the strong Np $f$-electron correlations, provide a realistic material-specific description of the electronic structure in the ordered phase, including the role and influence of each individual multipole. In particular, two specific properties of the LDA$+U$ framework make it suitable to study complex multipolar-ordered phases: the spin-orbital dependence of the local potential, which is essential since appearing multipole order parameters
involve multiple spin and orbital degrees of freedom, and the ability to take into account non-collinearity of local order parameters,
because the multipole moments on each Np site can be expressed through the local potential only. 

In the one-electron scheme, the strong spin-orbit coupling splits the $5f$ orbitals of a Np$^{4+}$ ($5f^3$) ion in  a $j$=5/2 sextet and a $j$=7/2 octet. NpO$_2$ crystallizes in the cubic CaF$_2$ ($Fm\overline{3}m$) structure, which consists of interpenetrating Np and oxygen FCC lattices. Under a cubic crystal field the lower $j$=5/2 multiplet splits into a $\Gamma_8$ quartet and a $\Gamma_7$ doublet. In the transition to the multipolar ordered state the atomic distances in the crystal structure are unchanged, yet the Np site-local cubic $O_h$ symmetry is broken.
The (111) orientation of the 3$\boldsymbol{q}$-antiferro electric quadrupoles,
detected in resonant x-ray scattering (RXS) experiments \cite{paixao02}, implies a lowering of the structural symmetry
to that of a 4-NpO$_2$ sublattices SC unit cell, where the Np site has local $D_{3d}$ symmetry \cite{sakai05}.
As a result, the $j$=5/2 states are split in two doublets, $\Gamma_6^{(1)}$ and $\Gamma_6^{(2)}$, and two singlets, $\Gamma_4$ and $\Gamma_5$; the latter two are actually degenerate if time-reversal symmetry is not broken \cite{dimatteo07}. 
Our selfconsistent calculations were performed using the full-potential linearized augmented plane wave (FLAPW) method, as implemented in the TSPACE and {KANSAI} program packages, combined with the LDA+$U$ approach \cite{suzuki10} to treat properly the on-site Coulomb correlations of the Np $f$ electrons. The relativistic spin-orbit interaction as well as the local +$U$ potential were included via a second variational step. For the LDA+$U$ double counting term  the fully localized limit was adopted.
Full-potential LDA+$U$ was shown to describe an anisotropic $f$-state \cite{suzuki10}, which is especially important for the multipole ordered state. The above-mentioned unit cell and the local $D_{3d}$ symmetry at the Np sites, together with broken time-reversal symmetry to accommodate magnetic solutions, have been used. Our $D_{3d}$-irreducible basis states $\Gamma_6^{(2)}$ ($\Gamma_4$, $\Gamma_6^{(1)}$, $\Gamma_5$) are constructed from the $\Gamma_7$ ($\Gamma_8$) orbitals in $O_h$ symmetry.
For the on-site Coulomb $U$ and exchange $J$ we use values of 3--5 eV, respectively, 0--0.5 eV, which are in the commonly accepted range for light actinides.
The two-electron integrals of the Coulomb interaction of the $f$ electrons were expressed in terms of  effective Slater integrals $F_{\kappa}$ (0$\le$$\kappa$$\le$6),  which are associated with $F_0$=$U$, and,
$F_4$=$(41/297)F_2$, $F_6$=$(175/11583)F_2$, and $J$=$(286F_2$+$195F_4$+$250F_6)/6435$ \cite{calcs}.

\begin{figure}[tb]
\begin{center}
\includegraphics[width=1.0\linewidth]{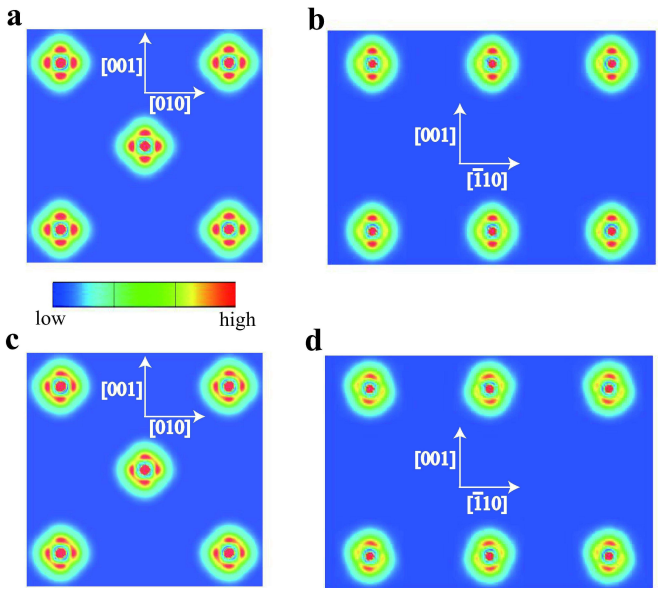}
\caption{(Color online) 
Computed charge density cross-sections of NpO$_2$.
{a,b,} Non-multipolar ordered state with cubic FCC symmetry in the [100], respectively, [110] plane.  
{c,d,} Magnetic multipolar state showing the (111) 3$\mbox{\boldmath$q$}$ antiferro electric long-range order in the [100], respectively, [110] plane.
Both solutions were obtained with $U$=4 eV and $J$=0 eV.
}
\label{Fig:cross-section}
 \end{center}
\end{figure}

Fig.\ \ref{Fig:cross-section} shows calculated cross-sections of the charge density of NpO$_2$
in the high-symmetry [100] and [110] planes. Panels {a} and {b} depict cross-sections for a non-multipolar ordered state, obtained for the FCC unit cell.
Panels {c} and {d} show cross-sections for a magnetic multipolar ordered phase, computed for the 4-sublattices SC unit cell, with the same values of the Coulomb $U$ and exchange $J$ parameters. A comparison of the charge densities in panels {a}, {c} and {b}, {d} reveals the symmetry reduction of the charge distribution in the multipolar ordered state, which is found to have a significantly deeper total energy than that of the non-magnetic single
FCC unit cell (in which multipolar order cannot occur). The 3$\mbox{\boldmath$q$}$-antiferro electric long-range order can be recognized clearly from panel {d}, where the charge lobes are tilted and point along ($\bar 1$11) directions.

\begin{figure}[tb]
\begin{center}
\includegraphics[width=1.0\linewidth]{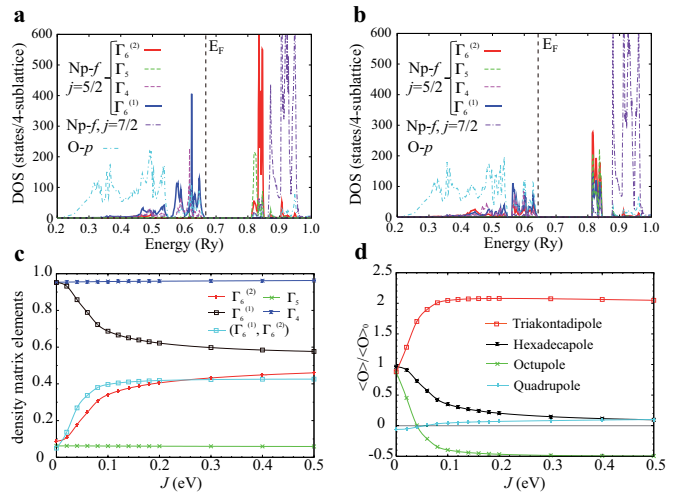}
\caption{ 
(Color online)
{a,} Orbital-projected density of states (DOS) of a multipolar state computed with $U$=4 eV,
$J$=0 eV.
Plotted is the Np $5f$, $j$=5/2, DOS projected on components of the irreducible $D_{3d}$ basis,
and the Np $j$=7/2 and 
O $p$ character DOS.
{b,} Orbital-projected DOS of the multipolar phase obtained with $U$=4 eV and $J$=0.2 eV.
{c,} Calculated LDA+$U$ one-electron orbital occupation numbers and off-diagonal ($\Gamma_6^{(1)},\Gamma_6^{(2)}$) matrix elements as a function of exchange $J$ at $U$=4 eV.
{d,} Calculated expectation values of the multipolar  (quadrupolar, octupolar, hexadecapolar, and triakontadipolar) order parameters in NpO$_2$ as a function of $J$.
}
\label{Fig:DOS-occ}
 \end{center}
\end{figure}

Fig.\ \ref{Fig:DOS-occ}{a}, {b} show the partial density of states (DOS) of the multipolar state computed with $U$=4 eV, $J$=0 and $U$=4 eV, $J$=0.2 eV, respectively. The Np $5f$, $j$=5/2, DOS is projected on components of the irreducible $D_{3d}$ basis.
The computed DOS reveals that NpO$_2$ in the multipolar ordered phase is an insulator with a gap of 2 eV  ($U$=4 eV, $J$=0 eV). 
In sharp contrast, cubic non-magnetic NpO$_2$ is computed to be metallic.
The orbital projected DOS panels {a} and {b} evidence a significant hybridization between oxygen $2p$ orbitals and Np $5f$, $j$=5/2 orbitals. 
The tiny contribution of the Np $j$=7/2 states to the valence states provides a ground to neglect them as was assumed in mean-field models \cite{kubo05a}. 
Fig.\ \ref{Fig:DOS-occ}{c} shows
the calculated diagonal (occupation number) and non-diagonal ($\Gamma_6^{(1)},\Gamma_6^{(2)}$) density matrix element as a function of $J$, with $U$ fixed at 4 eV.
When $J$=0, the occupied Np $5f$ orbitals have strong $\Gamma_4$ and $\Gamma_6^{(1)}$ character whereas the unoccupied orbitals have $\Gamma_5$ and $\Gamma_6^{(2)}$ character.
As soon as a non-zero value of $J$ is taken into account, the orbital character changes due to a mixing of the $\Gamma_6^{(1)}$ and $\Gamma_6^{(2)}$ states. The occupation of the $\Gamma_6^{(2)}$ orbitals starts to increase significantly, at the expense of the $\Gamma_6^{(1)}$ orbitals, and a relatively large off-diagonal term between the two doublets is present, which saturates for $J$$\ge$0.15 eV.
In order to obtain microscopic insight into the mechanisms which drive the phase transition,
we calculated the expectation values for the different multipoles having $\Gamma_{5}$ symmetry and ranks from 2 to 5 (electric quadrupoles, magnetic octupoles, electric hexadecapoles, and magnetic triakontadipoles) as a function of the exchange $J$ (Fig.\ \ref{Fig:DOS-occ}{d}), using their tensor-operator expressions $O_{\gamma \gamma^{\prime}}^{\tau \ell}$ \cite{kusunose08,cricchio09}:
\begin{equation}
\langle O^{\tau \ell} \rangle = \sum_{ \boldsymbol{k} b \gamma \gamma'}
\langle \boldsymbol{k}b| \tau\ell\gamma \rangle O_{\gamma \gamma^{\prime}} \langle \tau\ell\gamma^{\prime} |   \boldsymbol{k}b \rangle
= \sum_{\gamma \gamma^{\prime}} O_{\gamma \gamma^{\prime}}^{\tau \ell} n_{\gamma^{\prime}
\gamma}^{\tau \ell}
\end{equation}
where $|{\mbox{\boldmath$k$}}b \rangle$ are the Bloch states, $\tau$ labels the atomic site, $\ell$ the angular momentum quantum number, and $\gamma$ ($\gamma^{\prime}$)  indexes combined spin and orbital degrees. The latter can be expressed in an $\ell, m, s$ basis, or via a unitary transformation, in the irreducible basis.
 $n^{\tau\ell}_{\gamma\gamma'}$ is the 
density matrix in the muffin-tin sphere. 
Each multipolar moment has been normalized to the value which would be obtained if only the three lowest orbitals in the one-electron description, i.e., the $\Gamma_4$ singlet and $\Gamma_6^{(1)}$ doublet, were filled.
Our {\it ab initio} study demonstrates that, while the occupation of the $j$=7/2 octet can be safely neglected, all the orbitals which arise the $j$=5/2 sextet largely contribute to the ordered phase, with the effect to boost the expectation values of the generally discarded higher-rank multipole moments.

\begin{figure}[tb]
\begin{center}
\includegraphics[width=1.0\linewidth]{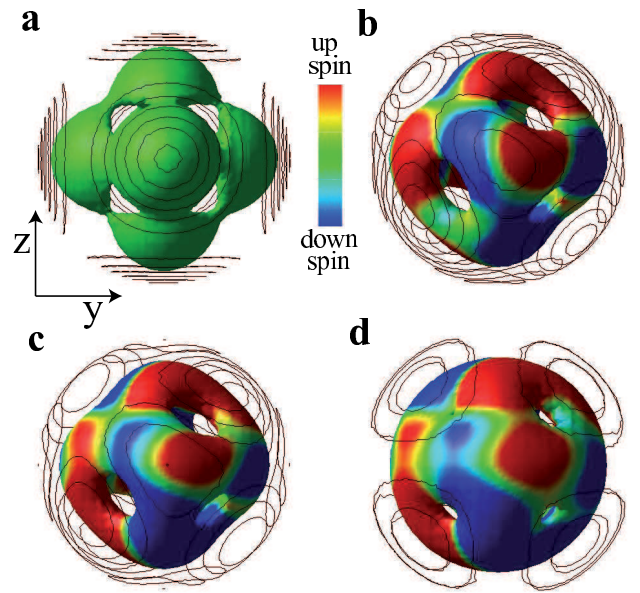}
\caption{(Color online) 
Computed charge and spin distributions of the Np
$5f$ electrons in NpO$_2$.
{a,} Non-multipolar ordered  state, computed 
with $U$=4 and $J$=0 eV.
{b,} Magnetic multipolar ordered state, for $U$=4 and $J$=0 eV.
{c,} Magnetic multipole state, for $U$=4 and $J$=0.04 eV.
{b,} Magnetic multipole state, for $U$=4 and $J$=0.2 eV.
The charge distribution is depicted by the three-dimensional hypersurface and the
direction of the current flow is shown by the thin lines. The colors on the three-dimensional charge distribution depict the local spin magnetization projected on the $z$ quantization axis. Red, blue colors label, respectively, a net up or down spin. Green color labels a vanishing net spin magnetization.
}
\label{Fig:3D-plot}
\end{center}
\end{figure}

We consider first the magnetic multipoles, whose the primary order parameters belongs. For $J$=0, their normalized expectation values coincide, an indication that triakontadipoles are inactive (and undistinguishable from octupoles). However, for non-zero $J$ values the ($\Gamma_6^{(1)},\Gamma_6^{(2)}$)
off-diagonal term increases. This mixing  ``unlocks'' the triakontadipolar degrees of freedom; whilst the expectation value of the octupoles is significantly reduced (changing its sign around $J$=0.04 eV and saturating at only about half of the original value for $J$$\ge$0.15 eV), the triakontadipolar moment is strongly enhanced. This finding is compatible with previous
model
calculations \cite{santini06} yielding a very low octupolar polarizability for the ordered ground state and unveils the microscopic mechanism which leads the $\Gamma_{5}$ triakontadipole to become the primary order parameter.

The effect observed on the electric multipoles as secondary order parameters is even more impressive.
The influence of the $\Gamma_6^{(1)}$-$\Gamma_6^{(2)}$ coupling
is so large for the quadrupole
that the normalized expectation values of the two electric multipole moments are different in magnitude and sign already when $J$=0 eV.
Increasing the $J$ value leads to a reduction of the hexadecapole and to a change in sign of the quadrupole,
but the former remains active and well-distinguished from the latter in the whole investigated 
$J$ range.
On these grounds, it might seem quite striking that the role of $\Gamma_5$ hexadecapolar degrees of freedom in the ordered phase of NpO$_2$ has never been considered
so far, but it must be reminded that hexadecapolar order could not be detected by a direct 
RXS experiment (in contrast to quadrupoles) due to the very small cross-section of $E2$-$E2$ transitions \cite{paixao02}.

Fig.\ \ref{Fig:3D-plot} shows calculated charge and magnetization distributions of the $5f$ electrons on a Np ion, for $U$=4 eV and several $J$ values.
Fig.\ \ref{Fig:3D-plot}{a} depicts the charge and spin distribution of the non-ordered state. 
The cubic $\Gamma_8$ symmetry of this distribution can be seen from the charge lobes pointing along the cubic axes of the CaF$_2$ structure.
The green color exemplifies that the distribution is non-magnetic everywhere. Panel \ref{Fig:3D-plot}{b} shows the distribution of the magnetic multipolar ordered state (for $J$=0 eV). The local symmetry axis of the charge distribution has become the (111) axis. Magnetic multipolar order is revealed by the pattern of the self-consistent spin distribution: a net up (down) magnetization along the $z$ axis is depicted by the red (blue) color. The dominantly red color occurring on the charge hypersurface around the (111) direction indicates a net up-magnetization along the (111) axis.
Fig.\ \ref{Fig:3D-plot}{c} shows the multipolar distribution computed for $J$=0.04 eV; an inspection of Fig.\ \ref{Fig:DOS-occ} shows that for this $J$ value the distribution contains no quadrupolar nor octupolar contribution, only a hexadecapolar and triakontadipolar contribution.
The fact that this multipole looks similar to the one in panel \ref{Fig:3D-plot}{b} allows one to infer that when the full $j$=5/2 multiplet is considered in selfconsistent calculations, in contrast with previous results obtained by simplified models, these high-rank contributions are  always important and at least as significant as lower-rank ones; in particular, the shape of the charge distribution in the small-exchange region is mainly due the electric hexadecapoles, whilst magnetic triakontadipoles dominate over the octupoles for all reasonable choices of $J$. Fig.\ \ref{Fig:3D-plot}{d} presents the obtained magnetic multipolar distribution for $J$=0.2 eV. This charge distribution maintains the symmetry along the (111) axis, but is more spherical. This can be understood as the result of a competition between the multipolar interaction and the crystal-field potential; the former tends to increase the occupancy of the $\Gamma_6^{(2)}$ doublet to the expense of states preferred by the latter, so that the cubic symmetry of the lattice is less evident (equal occupation of all states belonging to the whole $j$=5/2 sextet would automatically lead to spherical symmetry). The magnetic multipoles computed for larger $J$ values bear no significant difference from the one for $J$=0.2 eV.

Our state-of-the-art calculations 
offer new insight in the exceptional symmetry-broken phase of NpO$_2$.
The local symmetry breaking resulting from the ordering of both electric and magnetic multipoles, with the exception of dipoles, significantly lowers the total energy with respect to the non-magnetic state. Multipolar superexchange contributions drive NpO$_2$ to an insulating 3$\mbox{\boldmath$q$}$-ordered ground state.
The anomalous  phase originates from  a strong coupling of spin and orbital degrees caused by the relativistic spin-orbit interaction and  a competition between the crystal-field potential and the multipolar superexchange. 
Whilst the most common non-dipolar degrees of freedom (e.g. electric quadrupoles and magnetic octupoles) are present and active, 
we find that both the usually neglected higher-order multipoles (electric hexadecapoles and magnetic triakontadipoles) play at least an equally important role.
Our calculations emphasize the important influence of the higher-lying states in the multipole formation, and reveal that
a strong mixing between $\Gamma_6^{(1)}$ and $\Gamma_6^{(2)}$ states, due to an increased exchange $J$, causes a steep growth of the triakontadipole at the expense of the octupole.
Putting our first-principles results in context with the available experimental data, this allows to infer that the primary order parameter is the $\Gamma_{5}$ triakontadipole.

We thank R. Caciuffo and Y. Yun for valuable discussions.
This work has been supported by the Swedish Research Council (VR),
European Commission, SKB,
and Swedish National Infrastructure for Computing (SNIC).


\begin{thebibliography}{99}
\bibitem{calcs}
In the basis set we included the Np 5$f$, 6$d$, 7$s$, and 6$p$ orbitals as valence states and Np 5$d$, 6$s$ orbitals as semi-core states and for oxygen we treated the 2$s$ and 2$p$ states as valence states. The muffin-tin sphere radii were 1.4 {\AA} for Np and 0.9 {\AA} for O.
A plane-wave cut-off of 6.02\,$a^{*}$ was used, corresponding to 259 and 925 plane waves at the $\Gamma$ point for the FCC normal cell calculations and the 4-sublattice SC unit cell in the multipolar calculations, respectively. 
In reciprocal space we used for the selfconsistent convergence (charge density and DOS calculation) 12$\times$12$\times$12 (24$\times$24$\times$24) $k$-points  in the FCC unit cell and 6$\times$6$\times$6 (12$\times$12$\times$12) for the 4-sublattice SC unit cell.
\end{thebibliography}
\end{document}